\documentstyle[psfig]{article}

\begin{document}

\title{\sc AstroMD. A multi-dimensional data analysis tool for
 astrophysical simulations.} 
\author{U. {\bf Becciani}, V. {\bf Antonuccio-Delogu} \\
Osservatorio Astrofisico di Catania\\Via S. Sofia 78, I-95125 Catania - ITALY \\
C. {\bf Gheller}, L. {\bf Calori}, F. {\bf Buonomo}, S. {\bf Imboden}\\
Cineca\\Via Magnanelli 6/3, I-40033 Casalecchio di Reno (BO) - ITALY }
\maketitle
\date{}
\maketitle

{\it Keyword:} N-body Simulations, Analysis tool, Scientific Visualization

\begin{abstract}

Over the past few years, the role of visualization for scientific purpose  
has grown up enormously. 
Astronomy makes an extended use of  visualization
techniques to analyze data, and scientific visualization has became a fundamental 
part of  modern researches in Astronomy. With
the evolution of high performance computers, numerical simulations
have assumed a great role in the scientific investigation, allowing the user to 
run simulation with higher and higher resolution. Data produced in these
simulations are often multi-dimensional
arrays with several physical quantities. These data are very hard to manage
and to analyze efficiently. Consequently the data analysis and visualization tools 
must follow the new requirements of the research. AstroMD is a tool for
data analysis and visualization 
of astrophysical data and can manage different physical
quantities and multi-dimensional data sets. The tool uses virtual reality 
techniques by which the user 
has the impression of travelling through a 
computer-based multi-dimensional model. AstroMD will be a freely available 
tool for the whole astronomical community.

\end{abstract}

\section{\bf Introduction}

Since the beginning of modern astronomy, the scientific community
expressed a great interest in scientific visualisation tools. A strong boost
in this direction was determined by the introduction of modern CCD detectors to collect 
observational data in a digital form. Today, almost all the standard measures 
are digital and each measure can be generally considered
as a collection of images forming a multi-dimensional data set. 
In many cases extensive image 
processing is required to obtain meaningful images. Useful 
scientific information can be obtained from the raw data only using a data reduction pipeline. 
The most common data reduction tools, like IRAF \cite{iraf93} from National
Optical Astronomy Observatories (NOAO) and MIDAS \cite{eso95} from European Southern Observatory, make an extensive use 
of visualization. This is supported by popular tools like 
AVS, IRIS Explorer, Data Visualizer and IDL. However, these
are generic purpose packages, not specifically designed for astrophysics.
This makes often difficult  their usage and they often required
to learn a new specific programming language. Furthermore it 
is usually hard, and some time impossible, to add user-developed 
data analysis tools.

Besides observational data,
the improvement of technology and the availability of super-computing
multiprocessor system, have led to a dramatic increase of the volume 
of data coming from  numerical 
simulations. Astrophysical simulations produce today 
 gigabytes of data which have to be efficiently visualized and analyzed. 
 In fact a simulation
produces several output to be correlated and analyzed, corresponding
to different temporal tags and associated computational meshes 
(sometimes irregular or structured grids). Furthermore, particle methods \cite{hoc81} 
(N-body, SPH , etc) are, especially in cosmology, extremely 
popular.
For this class of data the use of the 
above mentioned tools is not adequate. Specific tools, like TIPSY \cite{tipsy},
have been created and are commonly used. However, these tools are quite limited
in their functionalities, cannot describe and handle mesh-defined fields, usually
associated to particles related quantities (like mass), and are not designed to
work efficiently with large amount of data. 

In this paper we present AstroMD, an analysis and visualization tool specifically
designed to deal with the visualization and  analysis of astrophysical 
data, avoiding or solving most of the previously described difficulties and 
limitations. AstroMD can manage different physical
quantities. It can find structures having a not well defined shape or symmetries, and  
performs quantitative calculations on a selected region or
structure. Furthermore, it  also makes use of virtual reality techniques which are particularly 
effective for understanding the three dimensional distribution of the fields, 
their geometry, topology or specific patterns.
The display of data gives the illusion of a surrounding medium into which the user is immersed. 
The result is that the user has the impression of travelling through a computer-based 
multi-dimensional model which could be  directly hand-manipulated. In this sense, 
the virtual reality is a progressive lowering of the 
barrier which separates  users from their data \cite{ear93}.

AstroMD is developed by the VISIT
(Visual Information Technology) laboratory at  CINECA (Casalecchio di Reno - Bologna) 
in collaboration with the Astrophysical Observatory of Catania (hereafter OACT).  
AstroMD is an open source completely free code which is freely available
 (see http://www.cineca.it/astromd). 

The plan of the paper is as follows. In section 2 we will describe the basic 
features of AstroMD and the Graphic User Interface, in section 3 we describe
the first test case on which AstroMD has been tested:
the data produced in cosmological 
simulations performed using the parallel N-body tree code of the OACT group.
The preliminary results and data manipulation performed using AstroMD on these data are presented in section 4. 
Finally, in section 5, we draw the conclusions and  future developments.

\section{The visualization tool: AstroMD}

Data produced by astrophysical simulations have 
peculiarities that make them different from data 
produced with other kind of simulation or experiment \cite{pe93}. The principal features can
be summarized as follows:

\noindent $\bullet$\quad Different physical species have to  be considered: for example,
cosmological simulations consider both baryonic and dark matter. 
These two components have different physical properties and they have to be treated
with different numerical approach. Dark matter is usually described by
N-body algorithms, while simulation carried out with baryons have a fluid-dynamical description (either
Eulerian or Lagrangian). Further components, like stars or different chemical
species, can be introduced and followed in a specific way. These kinds of data 
require different types of analysis  and different kind of visualization
techniques. Dark matter needs the analysis of  particle 
positions and velocity, 
while baryons require mesh based analysis and visualization. Furthermore particle associated
quantities, like the mass density or the gravitational potential, 
require their calculation and visualization on a mesh.

\noindent $\bullet$\quad Simulated structures in general 
have not a well defined shape
or particular symmetries. Furthermore they can be distributed with no regularity
in the computational box. Therefore it is necessary to have a clear 
3D representation and efficient and fast tools of navigation, selection, zoom
and the possibility of improving the resolution and the accuracy
of calculations in specific user-selected regions

\noindent $\bullet$\quad Evolution changes dynamically the properties of 
the simulated objects and the information that can be retrieved, therefore it is
important to control efficiently sequences of time-frames.

AstroMD is specifically designed to deal properly with all these
requirements. In the following subsections we will describe in detail
the basic features of the package, the visualization, the data
analysis, the graphic user interface and the  stereo-graphic
capabilities.

\subsection{VTK: a scalable visualization library}

AstroMD is developed using the  Visualization Toolkit (VTK) by Kitware,
a freely available software portable on several platforms which range from 
the PC to the most powerful visualization systems, with a good scalability.

VTK \cite{vtk99} is designed for 3D computer
graphics, image processing, and visualization. It includes a C++ class library and
several interpreted interface layers. VTK has
been ported on nearly every Unix-based platform (Irix, Solaris, Linux etc.)
and PC's (Windows NT and Windows 98).
The design and implementation of the library has been strongly influenced by
object-oriented principles.

The graphics model in VTK is at a higher level of abstraction than rendering 
libraries like OpenGL or PEX.
This means it is much easier to create useful graphics and visualization 
applications. In VTK applications can
be written directly in C++, Tcl, Java, or Python.
Using these languages it is possible to 
build powerful, fast and portable applications.

VTK supports a
wide variety of visualization algorithms including scalar, 
vector, tensor, texture, and volumetric methods and
advanced modelling techniques. It supports stereo-graphic rendering and
can be used for virtual reality visualization. 
Furthermore, being easily extensible, VTK allows the user to build ad hoc
implementation of specific data analysis modules.

\subsection{AstroMD basic functionalities}

The input data format presently accepted by AstroMD is the common unformatted
C standard. Binary quantities (e.g. vector components) are written
in a continuous sequence with no labels or other symbols in within. The same 
result can be obtained in Fortran using direct access files. However several data input formats like  
TIPSY and HDF will be soon integrated. AstroMD can use the full dynamic range 
of the data, in order to retain the highest 
   accuracy in the analysis.

Data are visualized with respect to a box which can describe the whole 
computational mesh
or just a fraction of it. A sub-box can be selected interactively
inside the parent box with a  
different spatial resolution, so that the user can focus on the most
interesting regions. Boxes can be translated, rotated, zoomed in and out
with respect to selected positions. Colors and luminosities can be chosen
freely by the user. Images of different evolutionary stages can be
combined in order to obtain a dynamic view of the behaviour of the system.

The program allows one to treat both particles and fields related data.
The distribution of the particles can be easily visualized by AstroMD using the 
particles positions, as shown in Fig. 1 and Fig. 2.\\

\begin{figure}
\psfig{file=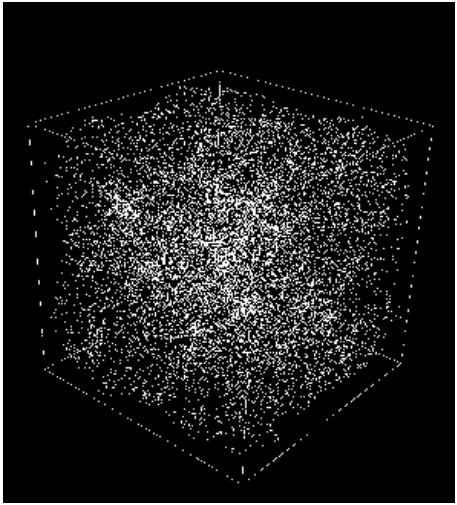,width=6cm}
\caption{AstroMD data visualization of a primordial particle distribution: 200,000 data points }
\end{figure}

\begin{figure}  
\psfig{file=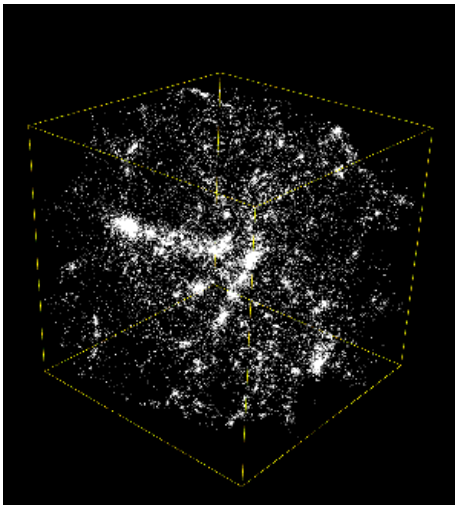,width=6cm}
\caption{Typical structure of voids and filaments at the end of a simulation}
\end{figure}

  The user can  choose to use a sample of the data, 
in order to get a faster and easier visualization. The opacity of the
particles can be increased, so that low density regions are more easily
detectable, or decreased, so that the details of the 
high density regions substructures are better visualized. Different particles species (e.g. dark 
matter and baryons) can be visualized at the same time using different
colours.
Other particle-related continuous quantities, like density fields or the
gravitational field, can be calculated as typical grid based fields.

In general, fields have a discrete representation on the cells of a 
computational mesh. The values of a field on a cell can be interpreted
as the average value of the field over that cell.  
In section 4 we will show several complementary ways of visualizing a field: different colours are associated to the
values of the analyzed field. 
Further possibilities of visualizing the field are either using cutting planes  where iso-value curves are drawn, or projecting the distribution on a plane,
integrating the field along the line of sight orthogonal to the plane. 
This last
representation, that is closer  to what actually comes from observations, 
will be soon included in the package.
Finally, different time frames can be shown in a sequence. When particles are used,
their positions are linearly interpolated between two available  key-frames. 
This technique engenders a graphic animation in real time, giving the impression of a continuous movement of particles, generally a 
randomized sample of the global data set. Both the single images and the 
whole sequence of time-step
can be saved in  bitmap format. Further output formats will be soon made available.

\subsection{AstroMD data analysis functionalities}

There are several AstroMD built-in tools which allow an
efficient manipulation and analysis of the data. 
The following functionalities are in progress and all of them will be 
integrated in  version 1 of the package.

\noindent $\bullet\ $ {\bf Particles mass density}

The mass density field associated to the particle distribution is 
calculated distributing the mass of each particle over the computational 
mesh by a eight points Cloud in Cell smoothing algorithm 
\cite{hoc81}. The computation 
can be done with the maximum accuracy using all the particles
over a uniform high resolution mesh, but AstroMD allows the user to use only a sample of the whole set of particles, and the final result can be extrapolated 
to all particles of the simulation, reducing the CPU time consuming and the memory request. 
The smoothing of the masses can be performed generally using a coarse grid, that
can be refined where high resolution is necessary.

The same tool can be used to calculate other fields related
to quantities possibly associated to the particles, like, for example, 
the thermal energy density field or the X-ray luminosity field.

\noindent $\bullet\ $ {\bf Gravitational field calculation}

Considering the mass density $\varrho(x)$  defined over the computational mesh
as above, the
gravitational field can be  calculated solving the Poisson equation
\begin{equation}
\phi(\vec x)\propto \nabla^2 \varrho(\vec x)\ ,
\end{equation}
where $\phi(\vec x)$ is the gravitational potential,
by a Fourier Transform procedure.
The  Poisson 
equation is transformed in its momentum space image using a FFT VTK
built-in function.
This reduces the equation to a much simpler algebraic operation
\begin{equation}
\phi(\vec k) \propto {1\over \vert \vec k \vert ^2} \varrho(\vec k)\ ,
\end{equation}
where $\phi(\vec k)$ and $\varrho(\vec k)$ are the Fourier images of the
potential and of the density and $\vert \vec k \vert ^2$ is the square
module of the wavenumber. Finally the potential is transformed
to the physical space using an inverse FFT.

\noindent $\bullet\ $ {\bf Fourier decomposition, power spectrum and correlation function}

The quantity $\varrho(\vec k)$ is used to calculate the power spectrum $P(k)$
of the matter distribution, which is defined as the average value
of the square norm of $\varrho(\vec k)$:
\begin{equation}
P(k) = \langle \vert \varrho(\vec k)\vert ^2  \rangle .
\end{equation}
The power spectrum expresses the weight of each of the Fourier components
of the mass distribution between $k_{min}$ and $k_{max}$ which 
represent the  inverse of the size of the computational
mesh and the Nyquist frequency. The power spectrum is a powerful 
measure of the statistical properties of the distribution, together
with the associated correlation function $\xi(r)$, which is its Fourier
transform. The correlation function indicates the probability to find 
a particle at a distance $r$ from any other particle, and is usually
used to analyze the clustering properties of a sample of discrete 
objects (particle, galaxies, galaxy clusters. etc.)

\subsection{The Graphic User Interface (GUI)}

The AstroMD Graphic User Interface has been projected for a simple 
and useful setting of the visualization parameters. \\
The commands are located in two main sections: a set of four menu buttons
on the top of the AstroMD window and two columns of commands on the
left side (Fig. 3). In the following we describe in detail the most important displayed commands.\\

\begin{figure}  
\psfig{file=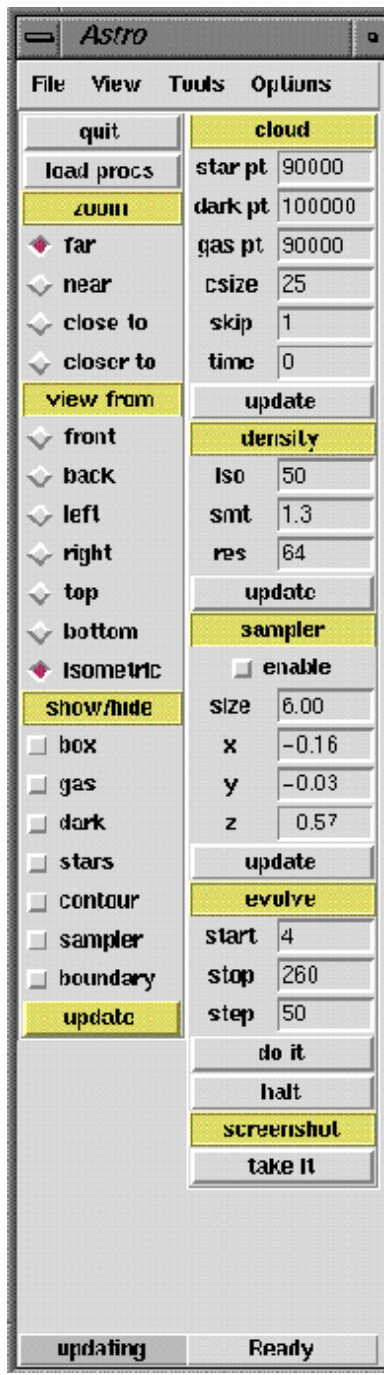,width=5cm}
\caption{The AstroMD GUI. The user can set all the parameters of the visualized fields}
\end{figure}

\paragraph{Menu Buttons.} 
The {\bf File} button contains the command to load three
types of data: it is possible to load data which can refer to
the gas, dark matter and stellar components  of a simulated system. The
different types of particles can be displayed all together or separately,
depending on the type of information the user wants to obtain.\\
The {\bf View} button allows the selection of a point of view
of the visualized system. These choices (being also reported in the  column
 of the commands), integrate the
possibility to rotate, scale and zoom into the domain of the visualized system, 
using the mouse.\\
With the {\bf Tools} button, the user can select several utilities. The
 {\it Interactor}  permits  to give VTK commands
interactively; the {\it Pipeline} shows the VTK Pipeline,
that is the network of filters which transform data into graphics
primitives and creates the visualized objects.

\begin{figure} 
\psfig{file=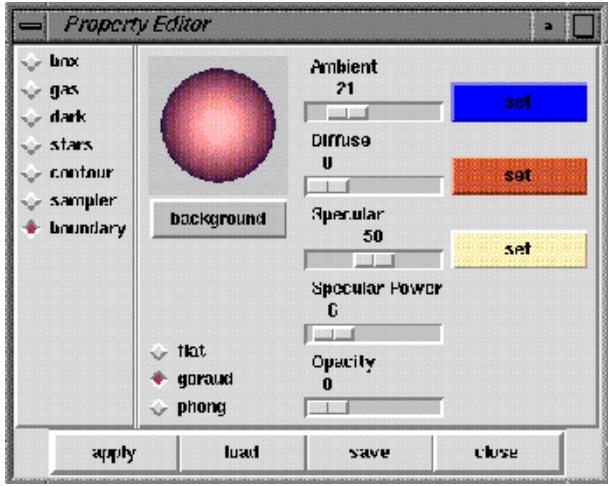,width=8cm}
\caption{Actor Properties to regulate the opacities of each visualized object}
\end{figure}

The {\it Opacity} selection allows to choice 
the values of opacity for the components of the astrophysical
system. A more accurate setting of the opacities can be performed using 
the {\it Actor Properties} (Fig. 4): for each 
visualized object, permits the user to regulate the
intensity and colour of the {\it Ambient light}, i.e. the light reflected or
scattered from other objects in the scene; the {\it Diffuse light}, which is
the light reflected by the object taking into account the angle of
incidence of the light onto the object itself, the {\it Specular light},
representing direct reflection of a light source off a shiny object, 
the {\it Specular power} which tells how the object reflects the incoming 
light and the {\it Opacity} which regulates the transparency of the
object: an opacity value of 0 produces the object to be completely
transparent. \\
The surface
colours of the displayed objects are obtained interpolating across 
polygons. The buttons, {\it Flat}, {\it Gourand} and {\it Phong}  correspond 
to three different types of interpolation: they are related to the
process of transformation from graphic primitives to the visualized objects.
The background colour can be selected  from a  button in the {\bf Tools} menu,
the combination of colours and lights of the objects can be saved and 
reloaded.\\
With the {\bf Options} button the user can select the stereo-graphic
modality of visualization, wide the rendering window up to full screen and
save the window geometry for particular situations.\\

\paragraph{Command Column.}
The {\bf Zoom} and {\bf View From} section of commands
allow  the regulation of the point of view of the objects while the {\bf
Show/Hide} section of buttons select the components of the simulated
system that the user wants to visualize. The {\it Box}, {\it Gas}, 
{\it Dark}  and {\it Stars} have obvious meanings; the {\it Contour} button
is referred to the visualization of iso-surfaces, the {\it Sampler}
activation permits the computation of the properties for a subset of
particles, and the {\it Boundary} button shows a periodic reproduction
 with one tenth of the total number of particles in the box including all the system.\\
The user can choose the sample of the loaded particle type 
(i.e. stars, dark matter and
gas particles), the size of the visualization box and the starting time from which
he wants to show the evolution of the simulation.\\
The {\bf Density} entries control the visualization of the iso-surfaces. These are 
calculated on a grid whose resolution can be user-selectable.\\
To allow the user to investigate with more accuracy a subset  of the visualized 
system it is possible to use a cubic {\it sampler}. If the sampler is
selected ({\bf Show/Hide} menu), visualized in the scene and enabled
({\bf Sampler} menu), all the 
computations are performed only inside the region of the sampler. \\
AstroMD can also show the evolution in time of the simulated system
over all the interval of time for which the data are available, performing
interpolations at intermediate frames. The {\bf Evolve} section permit to fix
the initial and final times of the evolution and the number of steps used
for the evolution. During the evolution the updated time of evolution is 
displayed  in the {\it Time} entry of the {\bf Cloud} section.\\
Finally snapshots of the images displayed can be created using the button
{\it Take it} in the {\bf Screenshot} menu. A subdirectory named 
{\it screenshot} is created (if it does not exist) and bitmap images (progressively numbered) are
created.\\
All the groups of commands can be changed and are applied  using the
corresponding {\it Update} command, the {\it Update} command at the end of
the left column updates all the system as a whole.

\subsection{Stereo-graphic visualization}

AstroMD can represent data using stereo-graphic visualization and display data
 in three dimensions with a stereo-scopic technique. The 3D stereo-scopic  perception, is based on the different angles we see a near object due to
the distance of the eyes. The combination of the two images determine
the position of the object in space. This process is reproduced 
by stereo-graphic visualization devices. The visualization system
displays the same image as it is seen from slightly different 
 points of view and presents it separately to each eye. This is done by 
different methods.
The most common is using shutter glasses (made with liquid crystals)
that act so that
each eye only sees the image intended. The final superposition 
of the two images gives the impression of a 3D scene
where th user is  full immersed.
The stereo images can be created on different kind of displays: the screen of
 a PC, the virtual theatre, up to the cave environments or
the head mounted displays

The stereo-graphic functionalities of AstroMD are developed in the
Virtual Theatre of CINECA. In this environment we can use two different
kind of displays: a wide cylindric screen ($9.4 \times 2.7$ m, $150^o$ view angle) and
a projective desk ($1.9 \times 1.3$ m, 29.1 dpi). The visualization is controlled 
by a SGI Onix2 Infinite Reality2 system with eight MIPS R10000 processors,
4 Gbytes DRAM and three graphic pipes.  

Stereo-graphic visualization gives a useful enhancement of common 2D rendering
for many applications. For example, the perception of the three dimensional
distribution of structures is much easier. Specific geometric or topological
features of the sample comes out clearly from the 3D immersive representation.
Furthermore, it is possible to analyze quantities not directly dependent on
the position, visualizing their correlations as planes or surfaces.
A typical example is the fundamental plane of elliptic galaxies, which relates
three observable quantities like the half-light radius, the mean surface brightness and
the central velocity dispersion of these systems.

\section{Numerical cosmology. Preliminary results and data manipulation}

In this section we will show some preliminary result we obtain applying the AstroMD tool to
the data obtained from a simulation of 16 millions of particles \cite{an99}  studying the formation and the evolution of the  Universe. The simulation has been performed using the N-body tree code
of the OACT group  run on the Cray T3E supercomputer of CINECA  \cite{bec97},
\cite{bec00}.
In the following subsection we briefly describe the tree code based on the 
Barnes-Hut algorithm (hereafter BH) \cite{barh86}, and the cosmological model we use to obtain the 
multi-dimensional data set. 
 
\subsection{Cosmological simulations}

Large cosmic structure simulations have improved enormously over the 
past decade both in terms of mass and force resolution. 
State-of-the-art N-body 
codes hardly allow one to deal with a number of particle N$ \geq 10^{7}$.\\
The cosmological N-body simulations aim at finding a
solution to a set of general relativistic equations of motions
describing the motions of particles (e.g. the postulated {\it Dark
Matter} particles) in the Universe. These equations can be
written in the form:
\noindent
\begin{equation}
{\rm \bf p}=ma^{2}{\rm \bf \dot{x}}, \hspace*{1cm}
\frac{d{\rm \bf p}}{dt}=-m{\rm \bf \nabla}\phi 
\label{eq1}
\end{equation}
\begin{equation}
\nabla^2\phi=4\pi Ga^{2}\{\rho({\rm \bf x},t)-\bar{\rho}(t)\}
\label{eq2}
\end{equation}

\noindent where $m$ is the mass of a particle, $G$ is the gravitational constant, $a=a(t)$ 
is the time-varying expansion parameter (the scale factor of the Universe), $\phi=\phi({\rm \bf x},t)$ 
and $\rho({\rm \bf x},t)$ are 
 the gravitational potential and the local  density respectively at comoving coordinate  ${\rm \bf x}$,  
 and
$\bar{\rho}(t)$ is the mean mass density of the universe.

\subsection{The simulation code}

The  BH tree algorithm is a $NlogN$ procedure to compute the gravitational force 
through a hierarchical subdivision of the
computational domain into a set of cubic nested regions that form an
adaptive tree data structure.
Generally speaking, the {\it bodies} evolve in time according to the laws of
Newtonian physics:
\begin{equation}
\frac{d\vec{x_i}}{dt}= \sum_{j \neq i} - \frac{Gm_j\vec{d_{ij}}}{\mid d_{ij}\mid^3}
\label{eq3}
\end{equation}
where  $\vec{d_{ij}}= \vec{x_i}-\vec{x_j}$.\\
The fundamental approximation of the tree codes consist in the approximation 
of the force component for the {\it i} body. Considering a region $\gamma$ the force component 
on {\it i} may be computed as:

\begin{equation}
\sum_{j\in\gamma} - \frac{Gm_j\vec{d_{ij}}}{\mid d_{ij}\mid^3} \approx \frac{GM\vec{d_{i,cm}}}
{\mid d_{i,cm}\mid^3} + \; \hbox{higher order multipoles terms}\; 
\label{eq4}
\end{equation}
 where $M = \sum_{j\in\gamma}m_j$ and $cm$ is the center of mass of $\gamma$.\\
The system dynamics involves the advancement of the trajectories of all 
particles, and this is carried out through a discrete integration of the trajectory
of each particle.\\

The code has been parallelized in order to be used on the CRAY T3E system using
both HPF-CRAFT and SHMEM libraries.

\subsection{The cosmological model and the data set}

As a test case for the AstroMD
tool we use the output of a  simulation with 16.777.216 particles run on a 
CRAY-T3E 256 processor based machine at the CINECA. The
system evolves in a cubic region with size of 50Mpc (about $50 \cdot 
10^{13}$ km). The Cold Dark Matter cosmological model was used, each particle
having about $ 1 \cdot 10^{9}$ solar masses. With this kind of cosmological model it is 
possible to study the formation 
and the evolution of galaxies and clusters of galaxies \cite{pe93}. 
The simulation starts at a time (red-shift) $z=50$,  where the latter parameter is a
standard measure of time adopted in cosmology.
The age of the Universe can be calculated from
the red-shift parameter $z$ using the following equation: $t=\frac{2}{3H_0}
\cdot (1+z)^{-\frac{3}{2}}$ where $H_{0}^{-1} =9.7776 \cdot 10^9 h^{-1} yrs.$, which is
valid for the cosmological models we have simulated.
At $z=50$  the age of the Universe was about 24 million  years, when the
matter of the Universe was still almost uniformly distributed. The simulation was stopped
at the value $z=0.005$, the 
actual age of the Universe.  The system evolve for about 270 time-steps using the 
opening parameter $\theta = 0.8$  and
clusters of galaxies start to form at about $z=10$. 
We collected 22 output check point files starting from $z=4$ up to the end of the 
simulation totalling 18 Gbytes of data (position and velocity for each particle). 
This files were managed and randomized so that each selection of the first N 
particles in the output file, correspond to a selection of N random particles 
in the box where the system evolves. The AstroMD use this randomized files and
the ancillary informations to  visualize data and to extract the scientific 
informations as specified in the following paragraphs.  

\section{Using AstroMD on simulated data}
  
Fig. 1 shows a primordial
distribution of dark matter particles inside the 50 Mpc 
cubic region of the  simulation above mentioned.
In the displayed picture, only a subset of 200,000 particles has been used, 
but owing to the fact that the data
are randomized, a very good representation of the distribution of particles
is obtained. This configuration is quite uniform, with small peaks of 
density. As it is well known,  during the evolution the dark  matter collapses 
onto these peaks
of density leading to the formation of at least three main clusters
of galaxies and several minor groups, embedded in a typical
structure of voids and filaments (Fig. 2).
The system evolution, from the primordial distribution of 
matter up to the formation of the final structures, can be visualized 
by AstroMD. It is possible to note  the augmentation of the velocity of collapse onto the 
proto-clusters and the oscillations of the matter density (around the main peaks)
during the formation of the clusters.
Using the mouse it is possible to rotate the box and analyze the shapes
of the clusters from different points of view; moreover, zooming into the
main structures, the details of the high density regions can be studied.
A detailed analysis  can be done setting a grid.
The Fig. 5  shows the matter density field: the iso-surfaces  are determined by connecting the cells in
 where the field has the same user-defined threshold value.
In this figure  the {\it Sampler} has been used 
to show the iso-surfaces around the main cluster of
galaxies in the simulation, fixing a value for the density 50 times greater 
than the mean background density, and using a grid of $64^3$.
The displayed iso-surface of the inner
distribution of matter has the ellipsoidal shape . In Fig. 5 the iso-surfaces are completely 
opaque but lowering the opacity parameter value it is possible to see the 
particles of dark matter inside the iso-surface.\\

\begin{figure} 
\psfig{file=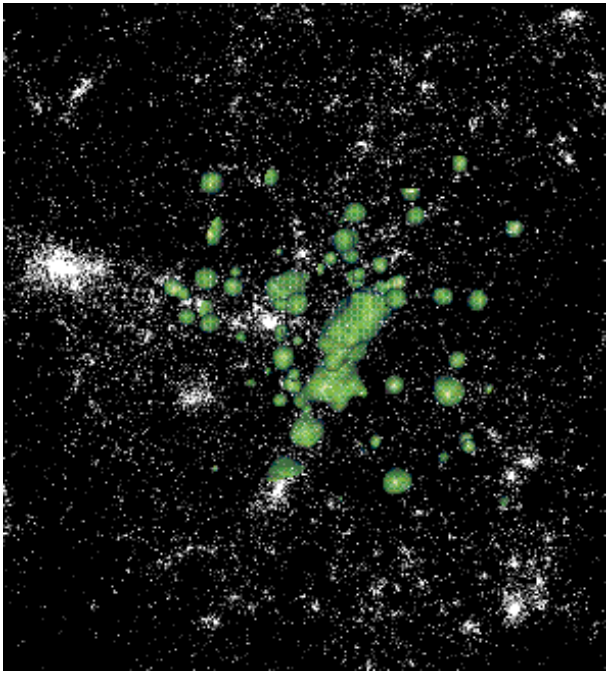,width=8cm}
\caption{Iso-surfaces, in a sub-region, at the end of a simulation.}
\end{figure}

Another different way of visualizing the field is by a volume rendering technique,
where the colour opacity  is a function of the field and its value 
typically increase with the magnitude of the field itself. 
Therefore, for example, 
low density regions appear more transparent (and differently coloured) than
higher 
density ones. Looking at the gas distribution in a cluster of galaxies,
we can see through the outer shells, until the inner, more dense and
more opaque core. In this way we mimic the actual behaviour of light
through a dense medium.

\section{Conclusions and future developments}

The enormous amount of data presently obtained both by observation and by 
numerical simulations, determines a growing request by astrophysicists of
powerful tools which combine efficient visualization and accurate analysis
capabilities. In this way both a qualitative and quantitative analysis of
the data can be obtained and great benefits can be achieved. Visualization,
in fact, allows one to have a more intuitive approach to the 
data, to find out easily their overall properties and characteristics and
to focus on the most interesting of them.
Data analysis gives the quantitative results required by the scientific
analysis.

We have presented AstroMD, a new data analysis and visualization tool,
which should respond to this demand. In the preliminary version
we have focused on the visualization aspects, with particular
attention to 3D immersive rendering and on 
creating a friendly graphic user interface. This makes AstroMD 
a useful instrument already competitive with  other existing software 
used in astronomy. The package is 
rapidly evolving and new functionalities, like a cluster identification
and reconstruction algorithm, are being added.
AstroMD has 
been built such that it is open source and  portable on all
the most common Unix systems and MS Windows. One of our basic purpose is to 
make it a wide spread product, highly diffused and used by the astrophysical
community.

\section {Acknowledgements}  

The immersive virtual reality was tested using the Virtual Theatre immersive visualization facility made available from CINECA. This is made up by: a graphic supercomputers, (SGI Onix2 Infinite Reality 2, 8 processors R10000, 4 Gbytes RAM); a high resolution projection system (3 projectors Barcographics); a cylindric screen (2.7x9.1 m, 2816x768 pixels); a projective desk (1.6x1.8 m, 29.6 dpi). 
We gratefully acknowledge useful discussion  with Dr. S. Bassini,
and we would like to thank G. Erbacci of CINECA for the useful help they gave us.

{}

\end{document}